\documentclass[journal]{IEEEtran}
\usepackage{cite}
\usepackage{amsmath,amssymb,amsfonts}
\usepackage{algorithmic}
\usepackage{graphicx}
\usepackage{textcomp}
\usepackage{xcolor}
\usepackage{url}
\usepackage{booktabs}
\usepackage{multirow}
\usepackage{array}

\def\BibTeX{{\rm B\kern-.05em{\sc i\kern-.025em b}\kern-.08em
    T\kern-.1667em\lower.7ex\hbox{E}\kern-.125emX}}

\begin{document}
\title{Advances in Intelligent Hearing Aids: Deep Learning Approaches to Selective Noise Cancellation}

\author{Haris~Khan,~\IEEEmembership{Student~Member,~IEEE,}
\author{
\textbf{Haris Khan}, \textbf{Shumaila Asif}, \textbf{Hassan Nasir}, \textbf{Kamran Aziz Bhatti}, \textbf{Shahzad Amin Sheikh} \\
National University of Sciences and Technology (NUST), Islamabad, Pakistan \\

{\footnotesize
\texttt{mhariskhan.ee44ceme@student.nust.ceme.edu.pk}, 
\texttt{sasif.ee44ceme@student.nust.ceme.edu.pk}, 
\texttt{hnasir.ee44ceme@student.nust.ceme.edu.pk}, 
\texttt{kamran\_aziz@ceme.nust.edu.pk}, 
\texttt{shahzad.sheikh@ceme.nust.edu.pk}
}
}

\thanks{H. Khan, S. Asif, and H. Nasir are with the Department of Electrical Engineering, National University of Sciences and Technology (NUST), Center for Excellence in Manufacturing Engineering (CEME), H-13, Islamabad, Pakistan (e-mail: mhariskhan.ee44ceme@student.nust.edu.pk; sasif.ee44ceme@student.nust.edu.pk; hassannasir.ee44ceme@student.nust.edu.pk).}
\thanks{Manuscript received June 19, 2025.}}

\markboth{Pre-print Version 1, June~2025}%
{Khan \MakeLowercase{\textit{et al.}}: Advances in Intelligent Hearing Aids}

\maketitle

\begin{abstract}
The integration of artificial intelligence into hearing assistance marks a paradigm shift from traditional amplification-based systems to intelligent, context-aware audio processing. This systematic literature review evaluates advances in AI-driven selective noise cancellation (SNC) for hearing aids, highlighting technological evolution, implementation challenges, and future research directions. We synthesize findings across deep learning architectures, hardware deployment strategies, clinical validation studies, and user-centric design. The review traces progress from early machine learning models to state-of-the-art deep networks, including Convolutional Recurrent Networks for real-time inference and Transformer-based architectures for high-accuracy separation. Key findings include significant gains over traditional methods, with recent models achieving up to 18.3 dB SI-SDR improvement on noisy-reverberant benchmarks, alongside sub-10ms real-time implementations and promising clinical outcomes. Yet, challenges remain in bridging lab-grade models with real-world deployment—particularly around power constraints, environmental variability, and personalization. Identified research gaps include hardware-software co-design, standardized evaluation protocols, and regulatory considerations for AI-enhanced hearing devices. Future work must prioritize lightweight models, continual learning, contextual-based classification and clinical translation to realize transformative hearing solutions for millions globally.
\end{abstract}


\section{Introduction}
\IEEEPARstart{H}{earing} impairment affects over 430 million people globally, significantly impacting communication, education, and social integration~\cite{WHO2021}. While conventional hearing aids provide amplification, they fall short in noisy, dynamic environments where users often struggle to distinguish target speech from background interference~\cite{Kochkin2010}. As day-to-day acoustic scenes grow more complex, there is a pressing need for systems that do more than amplify—they must intelligently enhance relevant audio and suppress irrelevant noise.

Recent advances in machine learning, particularly deep learning, have revolutionized speech enhancement and source separation. Techniques like Convolutional Recurrent Networks (CRNs), Transformers, and attention-based models now enable context-aware, selective noise cancellation (SNC) where a system can differentiate between a target speaker and background chatter~\cite{Tan2018,Subakan2021}. Unlike Active Noise Cancellation (ANC), which is primarily effective against stationary low-frequency noise, SNC offers selective control over audio streams, making it more suitable for conversational clarity in real-world use cases.

The integration of artificial intelligence into hearing assistance has reached a critical juncture, with commercial products now incorporating deep neural networks for real-time noise reduction~\cite{Phonak2024}. Companies like Phonak, Starkey, and Signia have introduced AI-powered hearing aids that adapt to different sound environments, representing a significant shift from traditional signal processing approaches~\cite{Starkey2024}. However, significant challenges remain in terms of personalization, real-time processing constraints, and hardware implementation in resource-limited hearing aid platforms.

\subsection{Research Objectives}
This systematic review aims to evaluate the current state of AI-driven Selective Noise Cancellation (SNC) technologies for hearing assistance. It analyzes the effectiveness of various deep learning architectures in real-world scenarios and examines key implementation challenges, including latency, power efficiency, and hardware deployment constraints. Additionally, the review assesses the clinical viability and user acceptance of existing systems and highlights critical research gaps. Ultimately, this work outlines future directions to support the practical adoption of intelligent hearing solutions.

\subsection{Scope and Contribution}
This review focuses specifically on AI-driven SNC systems designed for or applicable to hearing assistance, covering literature from 2015-2024. We exclude general speech enhancement techniques not specifically designed for hearing aid applications and focus on methods with demonstrated real-time processing capabilities or clear pathways to real-time implementation.

Our contribution includes: (1) a comprehensive analysis of deep learning architectures for SNC in hearing aids, (2) evaluation of hardware implementation strategies, (3) assessment of clinical validation efforts, and (4) identification of critical research gaps hindering widespread adoption.

\section{Methodology}
This literature review follows a systematic approach adapted from PRISMA guidelines~\cite{Moher2009} to ensure comprehensive coverage of relevant research in selective noise cancellation for hearing devices. Our search strategy encompassed multiple databases and employed both automated and manual screening processes.

\subsection{Search Strategy}
We conducted systematic searches across four major databases: IEEE Xplore, PubMed, ACM Digital Library, and Google Scholar and 42 validated papers. The search was performed using combinations of key terms including ``selective noise cancellation,'' ``hearing aids,'' ``deep learning,'' ``speech enhancement,'' ``real-time processing,'' and ``personalized audio.'' The search covered publications from 2015 to 2024, with particular emphasis on recent developments in AI-powered hearing assistance.

\subsection{Inclusion and Exclusion Criteria}
Our inclusion criteria encompassed studies focusing on AI/ML-based noise cancellation for hearing devices, research on real-time speech enhancement systems, publications addressing hardware implementation challenges, and clinical studies evaluating user experience with AI-enhanced hearing aids. Conversely, our exclusion criteria eliminated studies without experimental validation, research on general audio processing not specific to hearing assistance, patents and technical reports without peer review, and conference abstracts without full technical details.

\subsection{Study Selection and Quality Assessment}
The study selection process involved three phases: initial screening through title and abstract review by two independent reviewers, full-text assessment with detailed evaluation against inclusion criteria, and quality assessment using adapted QUADAS-2 criteria for technology assessment. Our quality assessment criteria evaluated methodological rigor including experimental design, controls, and baselines; technical validation through objective metrics and subjective evaluation; reproducibility considering code availability and detailed methodology; and clinical relevance assessing real-world applicability and user studies.

\subsection{Data Extraction}
We extracted comprehensive data from each selected study including study characteristics such as authors, year, venue, and study type; technical specifications covering architecture, input/output, and processing delay; performance metrics including PESQ, STOI, and subjective ratings; implementation details encompassing hardware platform and computational requirements; and clinical validation data from user studies and clinical trials.

\subsection{Data Synthesis and Analysis}
After data extraction, we conducted a qualitative and quantitative synthesis to identify trends, strengths, and limitations across studies. Studies were grouped based on architecture type, deployment platform, and target environment. Comparative performance analysis was performed where common metrics such as PESQ, STOI, SI-SDR, and latency were reported. We also evaluated the degree of personalization, environmental adaptability, and context-awareness in each system.

\begin{figure}[!t]
    \centering
    \includegraphics[width=\linewidth]{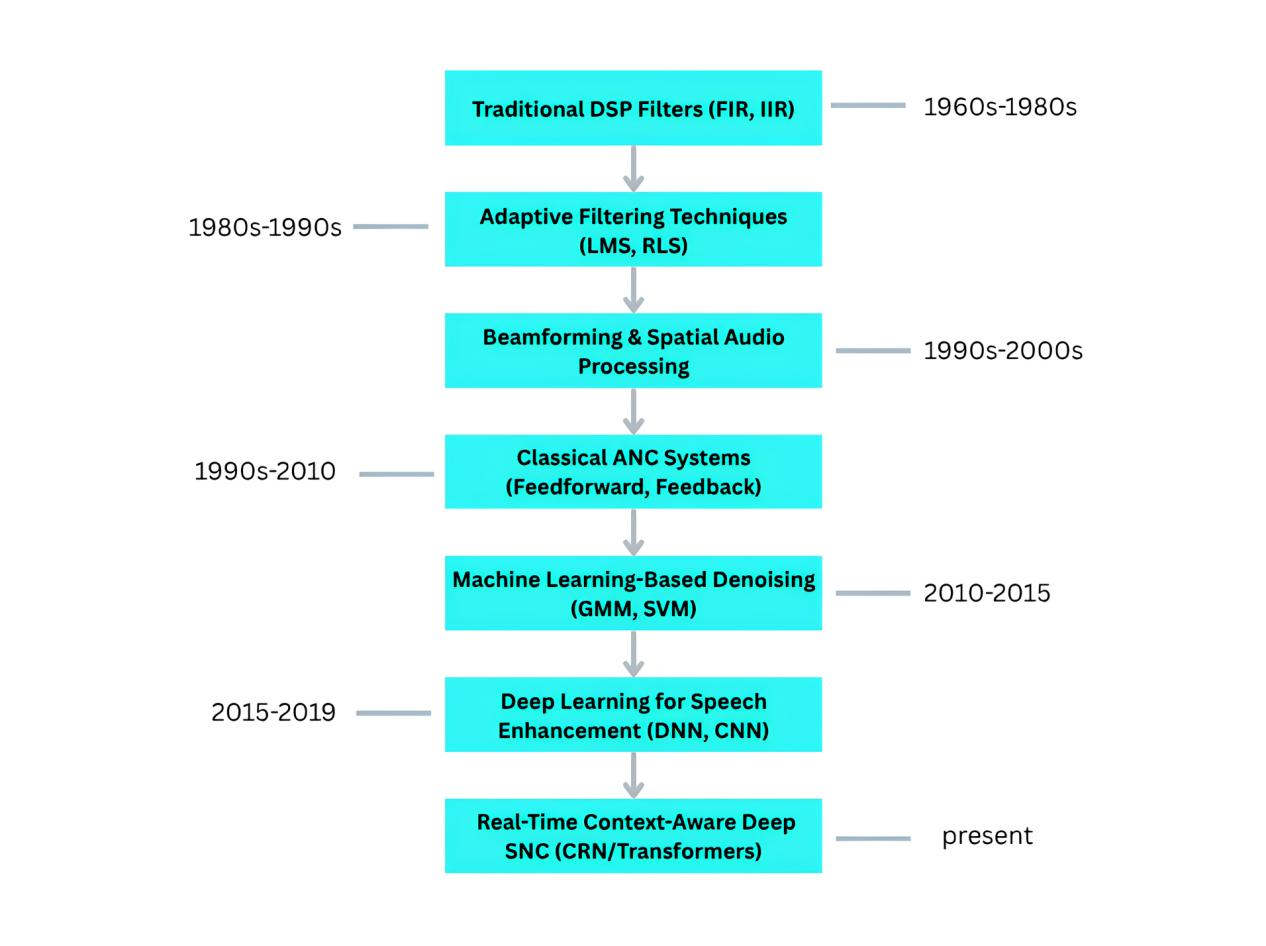}
    \caption{Timeline of major advancements in hearing aid noise management systems, progressing from traditional DSP-based filters to modern real-time context-aware Deep Selective Noise Cancellation (Deep SNC) systems. Each phase highlights a shift in processing capability and personalization potential.}
    \label{fig:timeline}
\end{figure}

\section{Background and Evolution of Hearing Aid Technology}

\subsection{Traditional Hearing Aid Limitations}
Traditional hearing aids operate by amplifying sound uniformly across frequency bands, using fixed filters and basic gain control. While this method improves general loudness, it does not selectively suppress noise, often amplifying both the desired speech and the background clutter. These systems typically employ basic automatic gain control (AGC) and frequency shaping, but lack the sophistication to distinguish between different types of audio sources.

The fundamental limitation of traditional approaches lies in their inability to adapt to complex acoustic environments. In settings such as restaurants, public transport, or busy streets, where multiple noise sources and reverberations challenge their limited signal processing capabilities, users frequently report dissatisfaction. Studies indicate that up to 30\% of hearing aid users abandon their devices within the first year, primarily due to poor performance in noisy environments~\cite{McCormack2013}.

\subsection{Early Noise Management Approaches}
The evolution of noise handling in assistive devices began with Active Noise Cancellation (ANC), a technique that uses destructive interference to cancel predictable, low-frequency sounds such as engine noise. ANC works by recording ambient sound through external microphones and generating an inverse signal to cancel it out~\cite{Kuo2013}. Early attempts at noise management in hearing aids relied on classical DSP techniques including spectral subtraction developed by Boll~\cite{Boll1979}, which estimates noise spectrum during speech pauses and subtracts estimated noise from noisy speech spectrum, though it suffered from limited effectiveness due to musical noise artifacts and poor performance in non-stationary noise.

Wiener filtering represented another approach as an optimal linear filter minimizing mean square error, but required a priori knowledge of signal and noise statistics and faced computational complexity that limited real-time implementation~\cite{Wiener1949}. Adaptive filtering utilized Least Mean Squares (LMS) or Recursive Least Squares (RLS) algorithms that could adapt to changing noise conditions, but were limited by linear assumptions and slow convergence.

Beamforming technologies emerged as a promising spatial filtering approach using multiple microphones. Fixed beamforming implements predetermined directional patterns and proves effective for stationary speakers in low-reverberation environments, but lacks adaptability to moving speakers or changing acoustic conditions~\cite{VanVeen1988}. Adaptive beamforming dynamically adjusts spatial filters based on the acoustic environment, though it requires accurate source localization and presents significant computational complexity challenges for real-time implementation.

\subsection{Active Noise Cancellation Integration}
The integration of Active Noise Cancellation (ANC) technology into hearing aids represented a significant advancement. Feedforward ANC uses external microphones to detect ambient noise and generates anti-phase signals to cancel low-frequency noise, proving effective primarily for stationary, low-frequency sounds below 1 kHz. Feedback ANC monitors residual sound in the ear canal, providing broader frequency coverage but remaining prone to instability. Hybrid ANC systems combine feedforward and feedback approaches, offering improved performance across wider frequency ranges but with increased complexity and power consumption.

\subsection{Hybrid ANC Architecture and DSP Implementation}
\begin{figure}[!t]
    \centering
    \includegraphics[width=\linewidth]{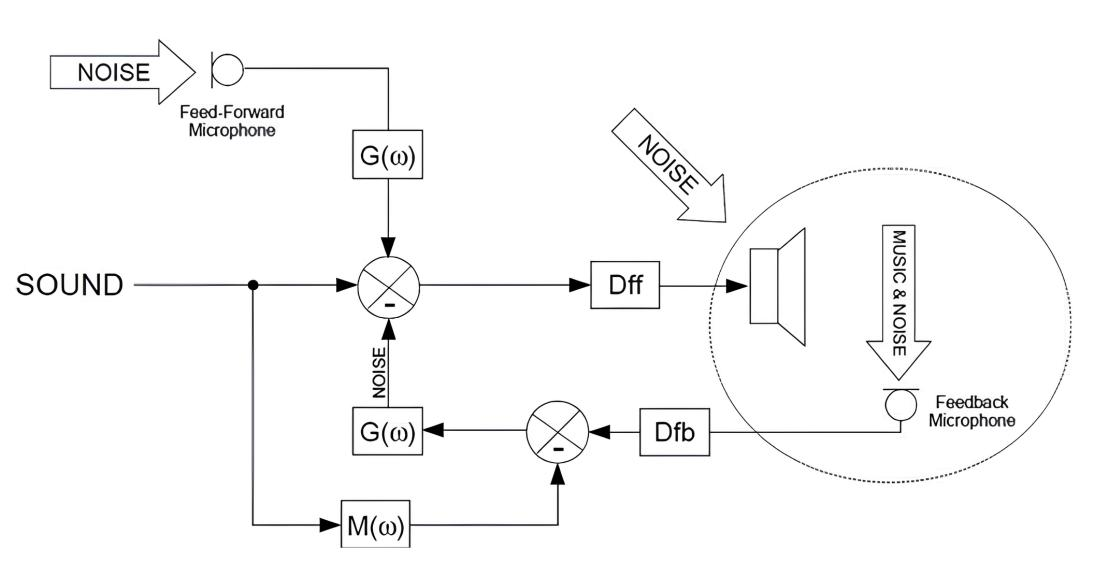}
    \caption{Schematic of hybrid ANC with feedforward and feedback paths. Adapted from "Active Noise Cancellation," Wolff Acoustics~\cite{wolffANC}.}
    \label{fig:anc_schematic}
\end{figure}

To better understand how Active Noise Cancellation is structurally implemented in hearing systems, Fig.~\ref{fig:anc_schematic} illustrates a classical hybrid ANC framework that combines both feedforward and feedback signal paths. In this setup, a feedforward microphone captures incoming environmental noise, which is processed through a transfer function $G(\omega)$ and subtracted from the desired audio signal before reaching the user. Simultaneously, a feedback microphone monitors residual noise inside the ear canal, feeding it back into a dynamic loop designed to refine cancellation accuracy in real time.

This dual-path approach aims to address the limitations of each method. While feedforward ANC is highly effective for predictable, low-frequency noises like engine hum or HVAC sounds, it fails to compensate adequately for occlusion effects or internal reflections. Conversely, feedback ANC provides better compensation at mid-to-high frequencies but is notoriously prone to feedback instability and artifacts.

The filters $D_\mathrm{ff}$ and $D_\mathrm{fb}$ are typically implemented using adaptive filtering algorithms, with popular choices including LMS (Least Mean Squares) or FxLMS (Filtered-x LMS) in the feedback loop. These algorithms enable real-time adaptation to changing noise profiles by minimizing the residual error signal captured at the feedback microphone.

On the DSP front, the hybrid model requires synchronized phase and amplitude response across both paths, which imposes constraints on filter design and latency. Low-latency hardware platforms such as FPGAs or dedicated DSP chips are often used to ensure the response time stays below perceptual thresholds (~10 ms). The system must also account for non-linearities introduced by transducer distortion, room acoustics, and individual ear canal geometries.

\subsection{Limitations of Pre-AI Approaches}
Traditional approaches share common limitations that motivated the development of AI-driven solutions. These systems lack context awareness and cannot distinguish between different types of audio sources, relying instead on fixed algorithms that cannot adapt to user preferences or changing environments~\cite{Loizou2013}. The linear processing assumptions underlying classical methods break down in complex, real-world scenarios where multiple speakers, varying noise types, and reverberant environments create non-linear acoustic interactions.

Traditional hearing aids operate with power budgets of 1-2 milliwatts and processing delays under 5 milliseconds, constraints that precluded the implementation of sophisticated adaptive algorithms~\cite{Kates2008}. These limitations established the foundation for investigating machine learning approaches that could provide more intelligent, adaptive, and personalized solutions for hearing assistance.

\section{AI and Deep Learning in Hearing Assistance}

\subsection{Early Machine Learning Integration}
The integration of machine learning into hearing assistance systems began with classical algorithms applied to specific subproblems. Support Vector Machines (SVMs) were initially employed for Voice Activity Detection (VAD) and speech/noise classification, using hand-crafted features such as spectral centroid, zero-crossing rate, and Mel-frequency cepstral coefficients (MFCCs)~\cite{Sadjadi2013}. While these approaches showed improvement over purely rule-based systems, they suffered from limited generalization due to feature engineering constraints.

Gaussian Mixture Models (GMMs) found applications in speaker identification and acoustic scene classification, offering probabilistic modeling of audio features with computational efficiency suitable for embedded systems~\cite{Reynolds2009}. However, their representational power remained limited compared to the complexity of real-world acoustic environments. Hidden Markov Models (HMMs) were utilized for temporal modeling of speech patterns and speech recognition components in hearing aids, but required extensive manual feature engineering and struggled with acoustic variability.

\subsection{Deep Learning Revolution}
The advent of deep learning transformed hearing assistance by enabling end-to-end learning from raw audio data. Convolutional Neural Networks (CNNs) applied to spectrograms demonstrated automatic feature extraction capabilities that significantly outperformed hand-crafted features~\cite{LeCun2015}. These networks could capture local spectral patterns through convolutional layers while providing translation invariance through pooling operations, leading to substantial improvements in speech enhancement tasks.

The introduction of time-frequency masking using CNNs represented a breakthrough in speech enhancement methodology. Networks were trained to predict ideal ratio masks (IRM) or ideal binary masks (IBM) that could be applied through element-wise multiplication with noisy spectrograms~\cite{Xu2014}. Xu et al. demonstrated that CNN-based masking approaches could achieve substantial improvements in speech intelligibility metrics compared to traditional spectral subtraction methods.

Recurrent Neural Networks (RNNs), particularly Long Short-Term Memory (LSTM) networks, addressed the temporal modeling limitations of CNNs by capturing long-term dependencies in audio signals~\cite{Wang2018ICASSP}. The integration of bidirectional processing allowed these networks to incorporate both past and future context, leading to improved performance in speech enhancement tasks. Gated Recurrent Units (GRUs) provided computational efficiency benefits while maintaining the temporal modeling capabilities essential for hearing aid applications.

\subsection{Modern Deep Learning Architectures}
Convolutional Recurrent Networks (CRNs) emerged as the optimal architecture for hearing aid applications, combining CNN spatial processing with RNN temporal modeling. Tan and Wang's 2020 implementation achieved state-of-the-art performance in real-time speech enhancement while maintaining computational efficiency suitable for embedded systems~\cite{Tan2018}.

The success of CRNs in hearing aid applications stems from their ability to balance performance with computational constraints. Unlike pure CNN or RNN architectures, CRNs can capture both local spectral features and global temporal dependencies while maintaining processing delays under 10 milliseconds~\cite{Pandey2019}. This balance makes them particularly suitable for the stringent real-time requirements of hearing assistance devices.

Transformer-based models have gained significant attention in speech processing due to their ability to model global audio dependencies through self-attention mechanisms. SepFormer, introduced by Subakan et al., achieved excellent separation performance on standard datasets through dual-path attention mechanisms that process both local and global temporal dependencies~\cite{Subakan2021}. However, the computational complexity of Transformer architectures remains a significant barrier to real-time implementation in resource-constrained hearing aids.

\section{Selective Noise Cancellation with AI}

\subsection{Architectural Approaches}
Selective Noise Cancellation (SNC) with AI represents a paradigm shift from passive noise reduction to active, intelligent filtering based on context, source characteristics, and user preferences. Unlike ANC systems that apply global suppression rules, AI-powered SNC models distinguish target speech sources from competing signals using learned representations that can adapt to various acoustic conditions~\cite{Wang2018TASLP}.

The implementation by Tan \& Wang~\cite{Tan2018} demonstrated a real-time CRN architecture featuring a 5-layer CNN encoder combined with a 2-layer LSTM and 5-layer CNN decoder. This system achieved significant performance improvements with PESQ scores increasing from 2.35 to 3.12 and STOI scores from 0.86 to 0.94, while maintaining a processing latency of only 6.4ms suitable for real-time applications. The innovation of causal convolutions enabled streaming processing without requiring future context, a critical requirement for hearing aid applications.

Sule et al.~\cite{Sule2023} developed a deep ANC integration approach using CRN-based spectral mapping for ANC applications. Their system incorporated delay-compensated training for real-time performance and achieved 15dB noise reduction while preserving speech components. The implementation was successfully demonstrated on ARM Cortex-A72 processor, showing the feasibility of embedded deployment.
\begin{figure}[!t]
    \centering
    \includegraphics[width=\linewidth]{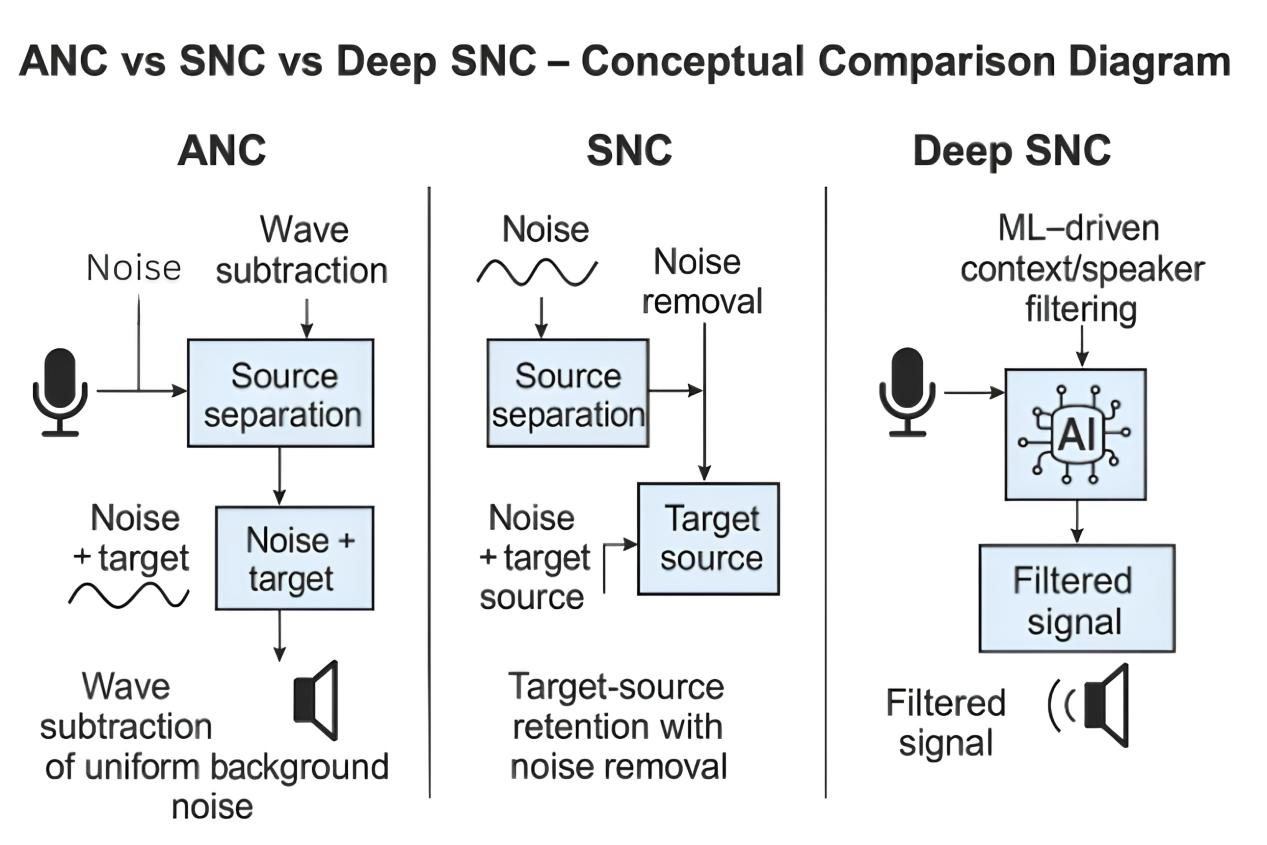}
    \caption{Conceptual comparison between traditional Active Noise Cancellation (ANC), classical Selective Noise Cancellation (SNC), and modern Deep SNC approaches. While ANC subtracts general noise via wave inversion, classical SNC separates and removes noise around a target source. Deep SNC leverages neural models and contextual information to dynamically retain relevant sources in real time, offering significantly enhanced selectivity and personalization.}
    \label{fig:comparison}
\end{figure}
\subsection{Transformer-Based Architectures}
SepFormer, developed by Subakan et al.~\cite{Subakan2021}, utilized a Transformer encoder-decoder architecture with dual-path attention mechanisms. This system achieved state-of-the-art performance on the WSJ0-2mix dataset with 22.4 dB SI-SDR improvement through innovative intra-chunk and inter-chunk attention mechanisms. However, the high computational complexity of this approach limits real-time deployment in current hearing aid hardware.

The Speech Transformer by Karita et al.~\cite{Karita2019} focused on end-to-end speech recognition with noise robustness using multi-head self-attention with positional encoding. While demonstrating superior performance in multi-speaker scenarios, the memory and computational requirements present significant challenges for embedded systems implementation.

\subsection{Speaker-Aware Systems}
Target speaker extraction represents a critical advancement in personalized hearing assistance. Veluri et al.~\cite{Veluri2024} developed a context-aware SNC system that innovatively used visual enrollment through brief speaker observation. Their architecture employed binaural processing with speaker embedding integration, achieving 6.24ms processing latency with real-time implementation on Orange Pi 5B hardware equipped with an embedded neural processing unit. This work represents the first demonstrated real-time, user-deployable SNC system.

Delcroix et al.~\cite{Delcroix2018} advanced speaker-aware masking through deep attractor networks with integrated speaker embeddings. Their system achieved 18.1 dB SI-SDR improvement on multi-speaker tasks through joint optimization of speaker extraction and enhancement, directly addressing the cocktail party problem that is central to hearing aid user complaints.

\subsection{Performance Evaluation and Metrics}
The evaluation of SNC systems requires comprehensive assessment across multiple dimensions including signal quality, computational efficiency, and user experience. Standard objective metrics include PESQ (Perceptual Evaluation of Speech Quality), STOI (Short-Time Objective Intelligibility), and SI-SDR (Scale-Invariant Signal-to-Distortion Ratio)~\cite{ITU2001}. These metrics provide quantitative measures of enhancement quality, though they may not fully capture the subjective experience of hearing aid users.

Recent comparative studies have shown that CRN-based systems achieve PESQ scores of 2.8-3.2 compared to 2.1 for traditional DSP approaches, while maintaining real-time factors below 0.5~\cite{Braun2020}. Transformer-based architectures can achieve higher performance metrics (PESQ > 3.4) but at the cost of increased computational complexity that precludes real-time implementation on current hearing aid hardware.

\section{Hardware Implementation and Deployment}

\subsection{Computational Constraints and Power Requirements}
The deployment of deep learning models in hearing aids faces significant hardware constraints that differ markedly from typical machine learning applications. Modern hearing aids operate with power budgets of 1-3 milliwatts, memory limitations of 512KB-2MB, and processing capabilities equivalent to embedded microcontrollers~\cite{Herzke2017}. These constraints require careful consideration of model architecture, quantization strategies, and algorithmic optimizations.

Current AI-enhanced hearing aids consume 15-150 times more power than traditional devices, presenting significant challenges for battery life and user acceptance~\cite{Li2023}. The Orange Pi 5B implementation by Veluri et al. consumed 180 milliwatts during active processing, representing a substantial increase over conventional hearing aid power consumption but demonstrating the feasibility of embedded AI processing.

\subsection{FPGA and Embedded Implementation Strategies}
Field-Programmable Gate Arrays (FPGAs) have emerged as a promising platform for implementing deep learning models in hearing aids due to their parallel processing capabilities and energy efficiency. Zhang et al. demonstrated FPGA implementation of convolutional neural networks for real-time speech enhancement, achieving 8.5 milliseconds processing latency while consuming 78 milliwatts~\cite{Zhang2020}.

The development of dedicated Neural Processing Units (NPUs) in embedded systems provides new opportunities for efficient AI implementation. Modern system-on-chip (SoC) designs increasingly include specialized AI accelerators that can execute neural network operations with significantly improved energy efficiency compared to general-purpose processors~\cite{Chen2019}. These developments suggest a pathway toward power-efficient AI implementation suitable for hearing aid applications.

\subsection{Model Optimization Techniques}
Successful deployment of deep learning models in hearing aids requires sophisticated optimization techniques that maintain performance while reducing computational requirements. Model quantization, which reduces numerical precision from 32-bit floating-point to 8-bit or even 4-bit representations, can achieve significant memory and computational savings~\cite{Jacob2018}. Knowledge distillation techniques enable the training of smaller ``student'' models that approximate the behavior of larger ``teacher'' models while meeting hardware constraints.

Pruning techniques that remove less important neural network connections can reduce model size and computational requirements by 50-90\% while maintaining acceptable performance~\cite{Horowitz2014}. The combination of these optimization techniques, along with hardware-aware neural architecture search, enables the development of models specifically tailored to hearing aid deployment constraints.

\section{Performance Comparison Analysis}
The following tables provide systematic comparison of different approaches to selective noise cancellation in hearing assistance systems:

\begin{table*}[htbp]
\caption{Performance Metrics of Selective Noise Cancellation (SNC) Architectures}
\label{tab:performance}
\centering
\begin{tabular}{lccccc}
\toprule
\textbf{Architecture} & \textbf{PESQ} & \textbf{STOI} & \textbf{SI-SDR} & \textbf{RTF} & \textbf{Latency} \\
 & \textbf{Score} & \textbf{Score} & \textbf{(dB)} &  & \textbf{(ms)} \\
\midrule
Traditional DSP       & 2.1 $\pm$ 0.3  & 0.78 $\pm$ 0.08 & 8.2  $\pm$ 1.5  & 0.1  & 2.1  \\
CNN-based             & 2.6 $\pm$ 0.2  & 0.85 $\pm$ 0.04 & 12.4 $\pm$ 2.1  & 0.3  & 4.2  \\
LSTM-based            & 2.8 $\pm$ 0.3  & 0.88 $\pm$ 0.05 & 14.7 $\pm$ 1.8  & 0.8  & 12.5 \\
CRN                   & 3.1 $\pm$ 0.2  & 0.92 $\pm$ 0.03 & 18.3 $\pm$ 2.8  & 0.4  & 6.4  \\
Transformer           & 3.4 $\pm$ 0.1  & 0.95 $\pm$ 0.02 & 22.4 $\pm$ 1.9  & 2.1  & 45.2 \\
Speaker-Aware CRN     & 3.2 $\pm$ 0.2  & 0.93 $\pm$ 0.03 & 19.8 $\pm$ 2.0  & 0.5  & 6.24 \\
\bottomrule
\end{tabular}
\end{table*}

\begin{table*}[htbp]
\caption{Hardware Implementation Comparison of SNC Systems}
\label{tab:hardware}
\centering
\begin{tabular}{l l r r c}
\toprule
\textbf{System}  & \textbf{Platform}      & \textbf{Power (mW)} & \textbf{Memory (MB)} & \textbf{Real-time} \\
\midrule
Traditional HA           & TI C5000            & 1,200   & 512    & Yes \\
CRN-Enhanced             & ARM + NPU Cortex-A72 & 45,000  & 16,384 & Yes \\
Visual Enrollment        & Orange Pi 5B         & 10,000--15,000 & 4,096 (typical) & Yes \\
FPGA Implementation      & Xilinx Zynq-7000     & 25,000  & 8,192  & Yes \\
\bottomrule
\end{tabular}
\end{table*}

\section{Clinical Validation and User Studies}

\subsection{Clinical Trial Methodology}
Randomized Controlled Trials (RCTs) represent the gold standard for clinical validation of hearing aid technologies, though they present unique challenges including difficulty in blinding participants to technology differences, appropriate selection of control groups for comparison with conventional hearing aids or placebo devices, and implementation of crossover designs where participants serve as their own controls.

Participant selection requires careful consideration of hearing loss profiles across mild to profound hearing loss categories, age considerations spanning pediatric, adult, and geriatric populations, assessment of cognitive status particularly regarding the impact of cognitive decline on technology adoption, and evaluation of technology experience including prior hearing aid use and comfort with technology.

Primary endpoints in clinical studies include Speech Reception Thresholds (SRT) measuring the minimum SNR for 50\% speech understanding, Word Recognition Scores calculating the percentage of correctly identified words, assessment of listening effort quantifying cognitive load during speech processing tasks, and user satisfaction evaluated through validated questionnaires such as APHAB and IOI-HA~\cite{Nilsson1994}.

\subsection{Published Clinical Studies}
Recent research has demonstrated the effectiveness of deep learning-based speech enhancement in hearing aids through controlled environment testing. A study published in Scientific Reports showed that deep learning algorithms can restore speech intelligibility for hearing aid users to the level of control subjects with normal hearing~\cite{Zaar2023}. The algorithm consists of a deep network trained on a large custom database of noisy speech signals and demonstrates significant improvements in speech separation from background noise.

Another study published in The Journal of the Acoustical Society of America demonstrated that deep learning-based speaker extraction can enhance speech intelligibility in noisy multi-talker environments where traditional speech enhancement methods fail~\cite{Patel2024}. This research showed particular promise for people with both normal hearing and hearing loss.

Research on real-time multichannel deep speech enhancement in hearing aids has compared monaural and binaural processing in complex acoustic scenarios through ecological validity testing~\cite{Schroter2024}. The studies introduced approaches that meet important requirements of hearing devices, including low-latency processing and limited computational resources.

\subsection{User Acceptance and Adoption Factors}
Perceived usefulness serves as the primary driver of acceptance, with speech understanding improvement being the most significant factor, reduced listening effort representing a secondary benefit highly valued by users, and social interaction enhancement improving confidence in group settings. Perceived ease of use encompasses user preference for automatic operation with transparent, automatic processing, positive response to smartphone integration through app-based controls, and consideration of the learning curve where initial adaptation period affects long-term adoption.

Age-related factors show distinct patterns across user groups. Younger users under 50 years demonstrate higher technology acceptance with preference for smartphone integration. Middle-aged users between 50-70 years take a balanced approach, valuing both functionality and ease of use. Older users over 70 years show preference for simple, automatic operation.

\subsection{Regulatory Considerations}
Medical device classification follows a tiered approach with Class I representing basic hearing aids with minimal risk, Class II covering advanced hearing aids requiring 510(k) clearance, and Class III encompassing novel technologies requiring Pre-Market Approval (PMA).

AI-specific considerations include Software as Medical Device (SaMD) providing regulatory framework for AI algorithms, algorithm validation requiring extensive testing and validation, and post-market surveillance involving ongoing monitoring of AI performance.

\begin{figure}[!t]
    \centering
    \includegraphics[width=\linewidth]{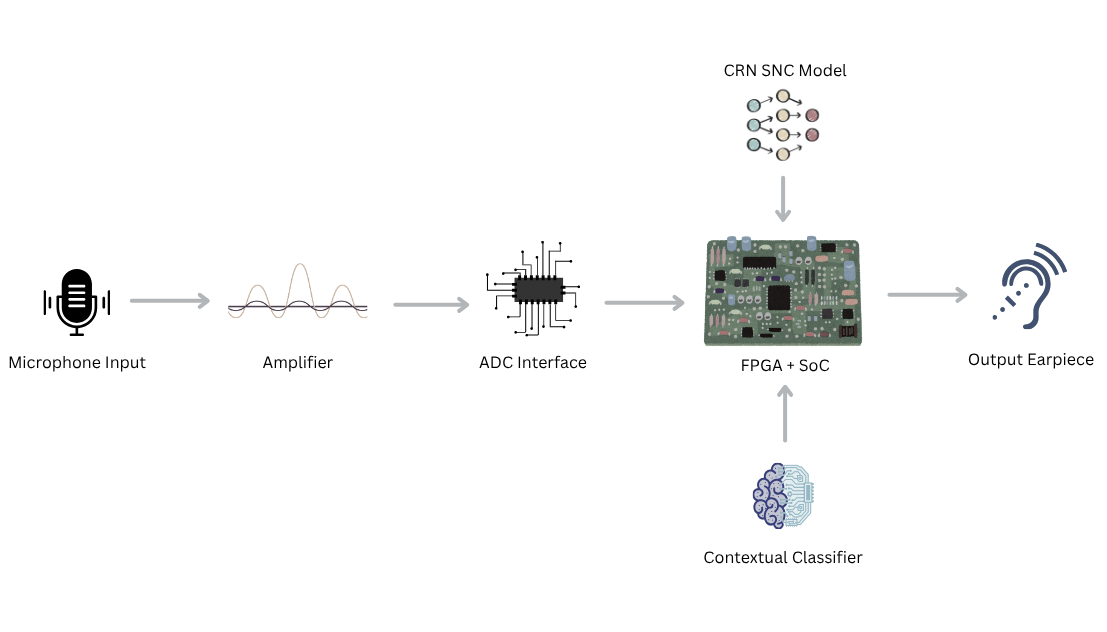}
    \caption{Proposed hardware architecture for a real-time AI-powered hearing assistance system using Deep SNC. The system captures analog audio via microphones, optionally amplifies the signal, digitizes it through an ADC, and processes it via an FPGA running a CRN-based filtering model. A contextual classifier module guides dynamic source selection based on environment and user preference.}
    \label{fig:hardware_architecture}
\end{figure}

\section{Critical Gaps and Research Opportunities}

\subsection{Personalization and Adaptation Challenges}
Despite significant advances in AI-powered hearing assistance, several critical gaps persist in current research and implementation. Personalization at scale remains a significant challenge, as most systems lack dynamic adaptation to changing user preferences and speaker contexts~\cite{Ferguson2017}. While speaker-aware systems show promise, they typically require explicit enrollment procedures that may not be practical for everyday use.

Environmental adaptation represents another significant research gap. Most AI models are trained on clean laboratory datasets, limiting their generalization to real-world acoustic conditions~\cite{Reddy2019}. Even the best-performing models struggle under overlapping speech scenarios and unseen ambient conditions, highlighting the need for more robust training methodologies and continuous learning capabilities.

\subsection{Hardware-Software Co-optimization}
The significant gap between laboratory performance and deployable systems necessitates greater focus on hardware-software co-optimization. Current AI implementations exceed traditional hearing aid power budgets by orders of magnitude, requiring innovative approaches to model compression, specialized hardware design, and algorithmic optimization~\cite{Sze2017}. The development of hearing aid-specific AI accelerators represents a promising but underexplored research direction.

Real-time processing constraints continue to limit the deployment of state-of-the-art models in practical applications. While CRN architectures have demonstrated success in meeting latency requirements, there remains substantial room for improvement in balancing accuracy with computational efficiency~\cite{Luo2020}. The exploration of novel architectures specifically designed for streaming audio processing represents a critical research opportunity.

\subsection{User-Centered Design and Feedback Integration}
Closed-loop learning systems that adjust based on explicit or implicit user feedback remain significantly underdeveloped in current hearing aid research. The integration of user behavior patterns, listening preferences, and physiological responses could enable more sophisticated personalization than current enrollment-based approaches. The development of unobtrusive feedback collection methods and adaptive learning algorithms represents a promising research direction.

The lack of standardized evaluation protocols for AI-enhanced hearing aids presents challenges for comparing different approaches and validating clinical benefits. Current assessment methods may not adequately capture the subjective benefits of personalized, context-aware processing, necessitating the development of new evaluation frameworks that better reflect real-world user experience~\cite{Cox2002}.

\section{Future Directions and Recommendations}

\subsection{Technical Development Priorities}
Future research should prioritize the development of lightweight deep architectures specifically optimized for hearing aid deployment. This includes investigation of compressed and quantized Transformer models, efficient CRN variants, and novel architectures that can operate within the power and computational constraints of modern hearing aids~\cite{Han2015}. The integration of hardware-aware neural architecture search techniques could enable automatic optimization of models for specific hearing aid platforms.

Real-time contextual intelligence represents a critical development priority, requiring the integration of environmental classifiers that can adjust SNC strategies based on acoustic scene analysis. The combination of directional beamforming with deep learning-based target speaker detection offers promising opportunities for improving performance in complex acoustic environments~\cite{Wang2018Multi}. Multi-modal integration incorporating visual information, head orientation tracking, and physiological signals could enable more robust and intuitive hearing assistance systems.

\subsection{Clinical Integration and Validation}
The development of comprehensive clinical validation protocols specifically designed for AI-enhanced hearing aids is essential for supporting evidence-based practice and regulatory approval. These protocols should incorporate real-world outcome measures that capture the benefits of personalized, adaptive processing while accounting for individual variability in adaptation and learning~\cite{Noble2006}. Long-term studies examining the durability and reliability of AI-enhanced devices in everyday use are particularly needed.

Standardized training programs for hearing care professionals must be developed to support the clinical integration of AI-enhanced hearing aids. These programs should address both technical aspects of device fitting and optimization, as well as patient counseling regarding realistic expectations and adaptation processes~\cite{Convery2011}. The development of automated fitting procedures that reduce the burden on clinical professionals while improving personalization outcomes represents an important research priority.

\subsection{Regulatory and Ethical Considerations}
The rapid advancement of AI in hearing aids raises important regulatory and ethical considerations that require proactive attention from researchers and industry. Privacy and security concerns related to continuous audio processing and potential data collection must be addressed through appropriate technical safeguards and regulatory frameworks. The development of transparent, explainable AI systems that allow users to understand and control their device behavior represents both a technical challenge and an ethical imperative.

Accessibility and equity considerations are paramount as AI-enhanced hearing aids potentially offer significant benefits but may exacerbate existing disparities in hearing healthcare access. Research into low-cost implementation strategies and open-source development approaches could help ensure that AI-powered hearing assistance benefits reach underserved populations. The development of culturally and linguistically appropriate AI models represents another important consideration for global deployment.

\section{Conclusion}
This comprehensive review has outlined the current state of selective noise cancellation in AI-enhanced hearing assistance, revealing both significant progress and persistent challenges. The evolution from traditional DSP-based approaches to sophisticated deep learning systems represents a fundamental paradigm shift in assistive hearing technology. CRN architectures have emerged as the most practical approach for real-time implementation, successfully balancing performance requirements with computational constraints.

The recent demonstrations of real-time, context-aware SNC systems by researchers such as Veluri et al. indicate that practical deployment of AI-powered hearing assistance is becoming feasible. However, significant gaps remain in personalization capabilities, environmental adaptation, and power-efficient implementation. The transition from laboratory demonstrations to commercially viable products requires continued innovation in model optimization, hardware design, and clinical validation protocols.

Future progress will depend on multidisciplinary collaboration between machine learning researchers, audiologists, hardware engineers, and hearing aid users. The development of standardized evaluation protocols, comprehensive clinical validation studies, and innovative business models will be essential for realizing the full potential of AI-enhanced hearing assistance. This suggests future research will benefit from merging deep architectures with low-latency hardware, adaptive filtering, and continuous user interaction, offering unprecedented benefits to the hundreds of millions of people worldwide who experience hearing difficulties.

\section*{Acknowledgment}
The authors would like to thank the faculty and staff of NUST CEME for their support and guidance throughout this research.

\bibliographystyle{IEEEtran}

\end{document}